\documentclass[11pt]{article}

\usepackage[final]{acl}

\usepackage{times}
\usepackage{latexsym}

\usepackage[T1]{fontenc}

\usepackage[utf8]{inputenc}

\usepackage{microtype}

\usepackage{inconsolata}

\usepackage{tabularx}
\usepackage{booktabs}
\usepackage{graphicx}
\usepackage{subcaption}
\usepackage{booktabs}
\usepackage{tcolorbox}
\usepackage{ragged2e}

\title{Representing data in words: A context engineering approach}

\author{Amandine M.Caut \thanks{Equal contribution}\\
  Information Technology Department\\ 
  Uppsala University \\
  \texttt{amandine.caut@math.uu.se} \\\And
  Amy Rouillard \footnotemark[1]\\
  Physics Department \\
  Stellenbosch University \\
  \texttt{rouillardamy@gmail.com} \\\And
  Beimnet Zenebe \footnotemark[1]\\
  Computer Science Department \\
  Addis Ababa University \\
  \texttt{beimnet.girma@aau.edu.et} \\
  \AND
  Matthias Green \\
  Insa Toulouse \\
  \texttt{mgreen@insa-toulouse.fr} \\\And
  Ágúst Pálmason \\
  Twelve Football \\
  \texttt{agust@twelve.football} \\\And
  David J.T.Sumpter \\
  Information Technology department\\
  Uppsala University \\
  \texttt{david.sumpter@it.uu.se}
  }

\begin{document}
\maketitle
\begin{abstract}
Large language models (LLMs) have demonstrated remarkable potential across a broad range of applications. However, producing reliable text that faithfully represents data remains a challenge. While prior work has shown that task-specific conditioning through in-context learning and knowledge augmentation can improve performance, LLMs continue to struggle with interpreting and reasoning about numerical data. To address this, we introduce wordalisations, a methodology for generating stylistically natural narratives from data. Much like how visualisations display numerical data in a way that is easy to digest, wordalisations abstract data insights into descriptive texts. To illustrate the method's versatility, we apply it to three application areas: scouting football players, personality tests, and international survey data. Due to the absence of standardized benchmarks for this specific task, we conduct LLM-as-a-judge and human-as-a-judge evaluations to assess accuracy across the three applications. We found that wordalisation produces engaging texts that accurately represent the data. We further describe best practice methods for open and transparent development of communication about data. 
\end{abstract}

\section{Introduction}\label{sec1}

\begin{figure}
    \begin{subfigure}{\linewidth}
    \centering 
    \includegraphics[width=0.85\linewidth]{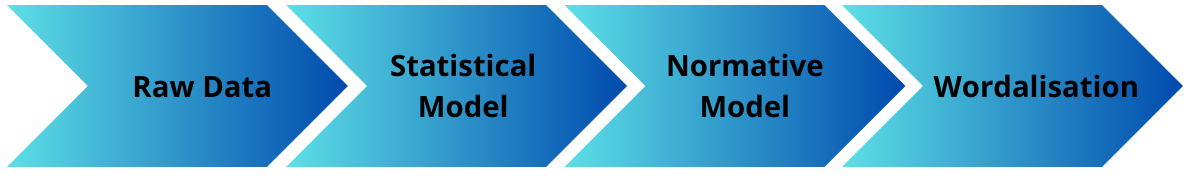}
    \caption{Workflow: A statistical model is applied to the raw data. The statistical description is converted into a synthetic text using a normative model. Combined with an appropriate prompt, the model generate a wordalisation, a high-quality, easy-to-digest text that faithfully represents the data. }
    \label{fig:wordalisation_workflow}
    \end{subfigure}
    \begin{subfigure}{\linewidth}
    \footnotesize
    \begin{tcolorbox}[colback=white, colframe=black, boxrule=0.5mm, arc=0mm, left=1mm, right=1mm, top=1mm, bottom=1mm]
    \textbf{Statistical model:} \\~\\
        \begin{tabular}{l r}
        \toprule
        \textbf{Factor} & \textbf{Z-score} \\
        \midrule
            Non-penalty expected goals & 2.101 \\
            Final third passes         & -0.663 \\
            Final third receptions     & -0.716 \\
            Ground duels               & -0.142 \\
            Air duels                  & -0.487 \\
            Smart passes               & -0.028 \\
            Goals                      & 2.041 \\
            Assists                    & -0.717 \\
            Key passes                 & -0.474 \\
            Minutes                    & 1.873
        \end{tabular}

    \end{tcolorbox}
    \begin{tcolorbox}[colback=white, colframe=black, boxrule=0.5mm, arc=0mm, left=1mm, right=1mm, top=1mm, bottom=1mm]
        \textbf{Normative model:}
        Here is a statistical description of [H. Kane], who played for [3201] minutes as a Forward. He was [outstanding] in [non-penalty expected goals] adjusted for possession per 90 minutes compared to other players in the same playing position. He was [below average] in [final third passes] adjusted for possession per 90 minutes compared to other players in the same playing position. He was [below average] in [final third receptions] adjusted for possession per 90 minutes compared to other players in the same playing position. He was [average] in [ground duels won] adjusted for possession per 90 minutes compared to other players in the same playing position. He was [average] in [air duels won] adjusted for possession per 90 minutes compared to other players in the same playing position. He was average in [smart passes] adjusted for possession per 90 minutes compared to other players in the same playing position. He was [outstanding] in [goals] adjusted for possession per 90 minutes compared to other players in the same playing position. He was [below average] in [assists] adjusted for possession per 90 minutes compared to other players in the same playing position. He was [average] in [key passes] adjusted for possession per 90 minutes compared to other players in the same playing position.
    \end{tcolorbox}
    \begin{tcolorbox}[colback=white, colframe=black, boxrule=0.5mm, arc=0mm, left=1mm, right=1mm, top=1mm, bottom=1mm]
        \textbf{Wordalisation:} 
       Harry Kane is an exceptional goal scorer who consistently creates high-quality scoring opportunities for himself. His outstanding goal and expected goals numbers highlight his clinical finishing ability and knack for finding dangerous positions. However, his contributions in creating chances for teammates are limited, as evidenced by his below-average assist and key pass numbers, and his involvement in the final third is also relatively low. Compared to other forwards, Kane's prolific goal-scoring ability makes him stand out significantly, despite the less-impressive numbers in other metrics. 
     \end{tcolorbox}
    \caption{Example wordalisation. The statistical model, here the z-score, of the football player Harry Kane is shown as a table (top). These are converted into a synthetic text using a normative model (middle). The normative model, in combination with the prompting template described in Section~\ref{sec:prompts}, is used to generate the example wordalisation for Harry Kane (bottom). } \label{fig:wordalisation_example}
    \end{subfigure}
   \caption{Workflow and example wordalisation.}
\end{figure}

Data science has become an important tool in business~\cite{davenport2020beyond}, biology~\cite{greener2022guide}, epidemiology~\cite{latif2020leveraging}, journalism~\cite{cairo2016truthful} and beyond. One way in which data science complements, and in some cases goes beyond, machine learning is in the use of visualisation to better understand datasets. Visualisation helps machine learning engineers both to understand their models and to communicate findings to non-technical users~\cite{zhang2020data}. Good visualisation practice involves understanding how to effectively communicate key facts about data without misleading the audience. For example, a series of articles by data journalist Lisa Charlotte Muth, and more recently Rose Mintzer-Sweeney, illustrate how careful design choices --- around colours, fonts, the placement of text and choice of chart --- are required when creating visualisations so as to convey information as clearly as possible~\cite{mintzer-sweeney_muth_2024}. Visualising data is not a neutral process, but rather an outcome of numerous decisions about how we want to present it to an audience~\cite{kennedy2016work}.  

Large language models (LLMs) currently demonstrate impressive mathematical and reasoning capabilities. However, despite their strong problem-solving strategies, they often struggle with presenting numerical data, partly because they treat numbers as text tokens rather than as quantitative information ~\cite{yang2025number, li2025exposing}.
In this article, we focus on communicating data in words using LLMs. Our aim is to "wordalise" data in a reliable way, going from raw data to a statistical model, to a normative model (i.e. an interpretation of the numerical value) and finally a wordalisation (see Fig.~\ref{fig:wordalisation_example} for an example of these steps). 
The wordalisation (a term we coin here) is generated by an LLM using a prompt constructed according to four task-agnostic steps, described in detail in Section~\ref{sec:prompts}, which build on in-context learning and few-shot prompting examples~\cite{brown2020language}. We have built a prototype application to demonstrate our idea, available at \url{ https://wordalisations.streamlit.app/}, which we encourage readers to visit to better understand our aims.

Unlike traditional data-to-text generation~\cite{lin2024survey,puduppully2022data,osuji2024systematic, wiseman2019learning} which aims to produce text from structured data, such as tables~\cite{Iso2019learning}, and rely on heavy training, our method are based on in-context learning using already pre-trained LLM to generate text. We differentiate our approach from other llm-based data-to-text generation schemes by addressing the semantic failing for such strategies~\cite{kasner2024traditionalbenchmarks}. Our work is under the umbrella of data-driven storytelling~\cite{schroder2023telling,shao2024data,zhao2023stories}, which aims to turn raw data into easy-to-read and easy-to-understand plain stories to help the user gain insights into data. In contrast to automatic captioning methods~\cite{you2016image, Hossain2019comprehensive, herdade2020image}, 
wordalisation aim to generate text directly from data, in an analogous way to how visualisation turns data into graphs. Visual elements can be used to support the understanding of the text, as shown in our demo application. 

Using prompt engineering to embed specific knowledge into LLMs~\cite{ahmed2024prompt,liu2025fewshot,song2025injecting} is quickly becoming the go-to method for improving answers provided by LLMs and counteracting hallucinations~\cite{chen2024unleashing}. Further, prompt engineering~\cite{amatriain2024prompt,schulhoff2024prompt,liu2023pre} can be used to shape the wordalisation generated for the specific user application. An advantage of using prompt engineering over model fine-tuning is that evaluating different prompts has a much lower computational cost,  performing as well and in come cases better than fine tuned models \cite{liu2022few}.  As Andrej Karpathy recently pointed out, what is needed goes further than prompt engineering, to encompass the idea of "context engineering" to provide language models with the right information, tools, and instructions appropriately formatted to enable reliable task completion ~\cite{context_engineering}.

\section{Application and Workflow}
\label{sec:applications}

 The originality of our work lies in the way we use prompt engineering applied to the LLM to convert raw numerical data into explanatory text that returns information consistent with the data itself, but also provides a judgment about the individual. The design aim of the wordalisation is to be both engaging to the reader and an accurate representation of the data source. Wordalisation can be applied across domains and to show this, we apply our framework to three application areas: football scouting, personality test analysis and international survey data exploration. Our method is applicable in scenarios where the user would like to compare a specific entity (single item in the dataset) to the larger population (the full dataset). We have built a Streamlit application~\cite{richards2023streamlit} to demonstrate our idea, and made it available here~\url{https://wordalisations.streamlit.app}.  Figure\ref{fig:wordalisation_example} gives examples of visualisations and corresponding wordalisation for each application area. 

 \begin{figure*}[h]
     \begin{subfigure}[b]{\columnwidth}
         \centering
         \includegraphics[width=0.7\linewidth]{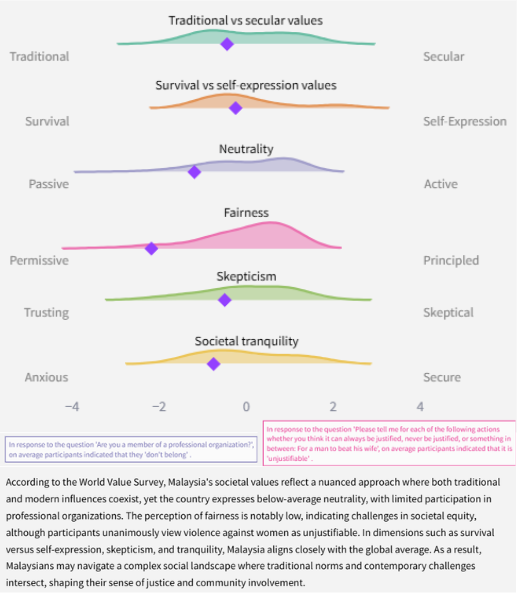}
         \caption{Data visualisation and wordalisation of the World Values Survey Wave 7: Case study of Malaysia.}
         \label{fig:top_image}
     \end{subfigure}
     \begin{subfigure}[b]{\columnwidth}
         \centering
         \includegraphics[width=0.85\linewidth]{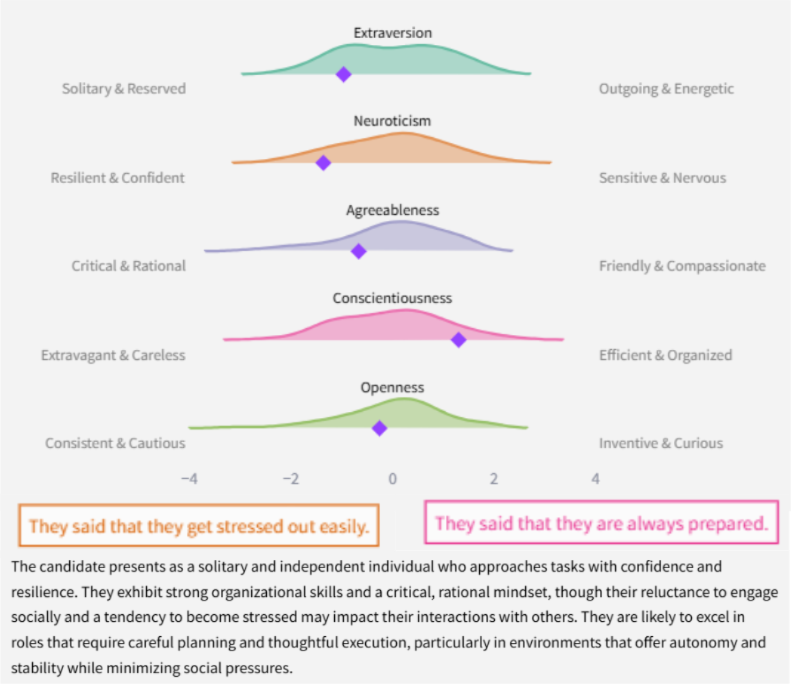}
         \caption{Data visualisation and wordalisation of the personality test: Case study of the person 69.}
         \label{fig:bottom_image}
     \end{subfigure}
     \caption{Data visualisation and wordalisation for psychometric and sociodemographic datasets: Applications to World Values Survey and personality assessment results.}
     \label{fig:data_visualisation}
\end{figure*}

Instead of simply telling the reader values or rankings, the wordalisation engages the reader. For example, by using familiar footballing terminology, while simultaneously accurately portraying the player's skills in shooting, passing the ball, competing defensively, heading and other skills an overview of the player is created. 
 
In the demo application, we also integrate the wordalisation with a chat functionality, using retrieval-augmented generation (RAG)~\cite{lewis2020retrieval}, allowing the user to ask questions about a specific individual or entity, in our case player, person, or country. 

The overall workflow for building  a wordalisation is illustrated in Fig.~\ref{fig:wordalisation_workflow}. Initially, informative statistics are calculated from the data, after which a normative model converts these numbers into a statistical description. This automated generation is done without the use of an LLM, using a template. As a result, it generates bland formulaic, but accurate texts. This synthetic text is then passed to a large language model as part of the prompt we outline below in order to generate the final wordalisation.

\section{Constructing the prompt}
\label{sec:prompts}

We now give a step-by-step account of how to construct the prompt underlying a wordalisation in four steps:
\begin{description}
    \item[Tell it who it is.] {The first step is to set the system prompt, which defines the role of the language model that will be used to generate text.}
    \item[Tell it what it knows.] {Here we give user-assistant prompts in the form of question-answer pairs that provide both information on how the language model should answer questions and provide background knowledge.}
    \item[Tell it what data to use.] {This step converts the content of the dataset and a selected data point into words. It allows for converting the statistical model into the normative model. }
    \item[Tell it how to answer.] {We provide examples of how the language model should write the final description from the text generated at the "tell it what data to use" stage.}
\end{description}

We now detail the 4 steps and give an example for each. By using the appropriate expander bar in the Streamlit app it is possible to see how the steps are implemented for each individual entity.

\subsection{Tell it who it is}

When calling an application programming interface (API) of a large language model, the system prompt tells the LLM the role the assistant should take when answering the questions. In the case of the football scout, we write:

\begin{tcolorbox}[colback=white, colframe=black, rounded corners]
"role": "system",\\
"content": ( "You are a UK-based football scout. You provide succinct and 
to the point explanations about football players using data. You use the 
information given to you from the data and answers to user/assistant pairs 
to give summaries of players.") 
\end{tcolorbox}

In generating a response the LLM pays special attention to this first system prompt, so it is important to tell it clearly what to do. For wordalisation we suggest being specific about the type of task that will be carried out, but to not yet give details of how it will be performed.

\subsection{Tell it what it knows}

After the system prompt we provide a sequence of question-answer (Q\&A) pairs, which help the LLM interpret the data it will be provided. These Q\&A pairs primarily illustrate how a domain expert would explain their knowledge to a non-expert. It is therefore essential that these answers both clarify what the data can tell us and highlight its limitations. These can be written by an expert and/or adapted from a reliable source. For example, for personality traits we used descriptions from Wikipedia of the big five personality traits~\cite{matthews2003personality} and for the football scout application we asked a domain expert to describe football statistics in terms of player performance.

\begin{table}[t]
\centering
\begin{tabularx}{\linewidth}{p{2.6cm} | X}
\multicolumn{1}{c|}{\textbf{Question}} & \multicolumn{1}{c}{\textbf{Answer}} \\
\hline
What is a forward? 
& A forward is a football player whose primary role is to score goals or assist teammates. They typically receive the ball inside the penalty area or make attacking runs high up the pitch. \\

What does it imply if a forward is good or excellent in terms of goals? 
& A forward who is good or excellent in terms of goals scores a lot. Goals are the most important thing in football, but we need to check that the player also does well in expected goals in order to check whether they might just have been lucky. 
\end{tabularx}
\caption{Example of a question–answer pair from the football scout scenario.}
\label{tab:football_example}
\end{table}

Constructing Q\&A pairs by hand ensures the relevance and accuracy of the Q\&A pairs. However, this task could also be semi-automated using an LLM. For such an approach, it is important to ensure that all phrases from the original text are accurately represented. Q\&A pairs should give insight into the data over and above simple definitions, and convey the desired language and tone.

\subsection{Tell it what data to use}\label{sec:tell_it_what_data_to_use}
Wordalisation assumes an underlying {\it statistical model} to effectively interpret data. In the case of the three applications presented here, we assume we can assess an individual (or entity) $i$ in terms of the z-score over one or more metrics, 
\[ z_{i,j} = \frac{x_{i,j} - \mu_j}{\sigma_j}, \]
where $\mu_j$ is the mean and $\sigma_j$ the standard deviation of the metric $j$. The z-scores are useful because they standardise the metrics and allow us to, for example, compare a player in terms of receiving and passing the ball on the same scale, defined by how far their z-score is from that of the average player in terms of standard deviations. Following statistical convention, one standard deviation can be considered to be larger than usual, two standard deviations can be considered to be significantly larger, and so on.  
The z-scores, in our case, are then formatted as a synthetic text, with phrases based on threshold ranges. For the football scout application, the threshold for z-scores are chosen as follows: above $1.5$ is described as "outstanding", between $1.0$ and $1.5$ is described as "excellent"; between $0.5$ and $1.0$ is described as "good"; between $-0.5$ and $0.5$ is described as "average"; between $-1.0$ and $-0.5$ is described as "below average"; and below $-1.0$ is described as "poor". We apply these thresholds to all the metrics for the player to generate a synthetic text. An example of this for Harry Kane is shown in Fig.~\ref{fig:wordalisation_example}. For illustration purposes, square brackets are used to highlight the selected phrases, and all square brackets are omitted in the final prompt. In these texts, we avoid presenting specific numbers by instead presenting a written interpretation of the data. We can think of this as a {\it normative model} for how we interpret the player's actions: more goals are better, as are more passes into the final third, etc. 
It is important to note that a wide range of statistical models can be applied here.

\subsection{Tell it how to answer}
The "tell it how to answer" step is a two-part process that involves providing the LLM with guidelines on response structure and incorporating a few-shot prompting technique. Firstly, the instruction part of this step can be used to detail instructions such as “answer in $3$ concise sentences” or “only use the data provided”. These directives serve as a precise framework, specifying the desired format and tone of the LLM’s responses.

\begin{tcolorbox}[colback=white, colframe=black, rounded corners]
"role": "user",\\
"content": ( Please use the statistical description enclosed with ``` to give a concise, 4 sentence summary of the player's playing style, strengths and weaknesses. "
"The first sentence should use varied language to give an overview of the player. "
"The second sentence should describe the player's specific strengths based on the metrics. "
"The third sentence should describe aspects in which the player is average and/or weak based on the statistics. "
"Finally, summarise exactly how the player compares to others in the same position.) 
\end{tcolorbox}

Secondly, few-shot prompting is used to demonstrate examples of how the LLM should adhere to the instructions. This involves including example pairs of statistical descriptions and corresponding assistant responses within the prompt. These pairs also reinforce the explicit directive given. We found this step to be essential, significantly improving the quality of the LLM's responses. These examples influence various aspects of the response, including structure, formatting, tone, tense, and which elements to emphasise. The LLM then closely adheres to these patterns when discussing similar data, ensuring consistency and clarity in the responses. 

\section{Evaluation and documentation} \label{sec:evaluation}
Assessing the performance of a large language model on a specialised task cannot be done via a standardised benchmark because each wordalisation application has its own unique and subjective set of goals and assessment criteria~\cite{raji2021ai, kaddour2023challengesapplicationslargelanguage, vendrow2024large}. Instead of benchmarking we propose a holistic approach for evaluating the performance using a combination of LLM-as-a-judge ~\cite{gu2024survey,li2024generation} and human-as-a-judge  methods with thoughtful documentation. 

\subsection{LLM-as-a-judge}
To measure the accuracy, we use Google’s "gemini-2.0-flash" to generate texts in three different ways: a {\it control} text, the LLM receives no information related to the entity. This allows us to assess whether the model already possesses prior knowledge about it. A {\it statistical} text we send only the raw statistics. We compare these two approaches to the {\it wordalisation}. The resulting texts from these three methods were evaluated by asking Google’s "gemini-2.0-flash": “Does the generated text contain implications or draw conclusions that are not supported by the statistical description?" which gave a binary classification of 'yes' or 'no'. To discourage the LLM from declining to respond when no description is given, we added the sentence "If no data is provided, answer anyway, using your prior statistical knowledge". 

 The control prompt generates factual inaccuracies in $91\%$ of cases, the numerical statistical description gave inaccuracies in $50\%$ of cases, while using the full Wordalisation texts and $18 \%$. Pairwise comparisons using z-tests for proportions confirmed that using the normative description of the data significantly increases accuracy in the text compared to both the control and numerical conditions ($p < 0.001$ in both cases).

 To investigate whether it is possible to accurately reconstruct the statistical description of a data point from the wordalisation we compare the statistical description that lies in the {\it synthetic text} to a {\it reconstructed synthetic text}. Given a `wordalisation', we instruct the LLM to classify each factor, according to the classes determined by the normative model, then to reconstruct the synthetic text. The original synthetic text serves as the ground truth for evaluation. The accuracy of the reconstructed statistical description serves as a measure of how faithfully the wordalisation represents the data point. For example, for the wordalisation of Harry Kane shown in Fig.~\ref{fig:wordalisation_example}, it is possible to (correctly) predict the class `outstanding' for the `goals' factor. We prompt the LLM to predict a classification for each relevant factor. We then compare the `true' class determined by data and the normative model with the predicted class. In cases where the generated json file could not be parsed, the data was discarded. To take into account random variations in the wordalisation due to the stochastic nature of the language model, we passed each prompt to the LLM multiple times, generating several wordalisations for each data point. 

\begin{figure*}[h!]
      \centering
      \includegraphics[width=0.75\linewidth, trim=0 0 0 1.4cm, clip]{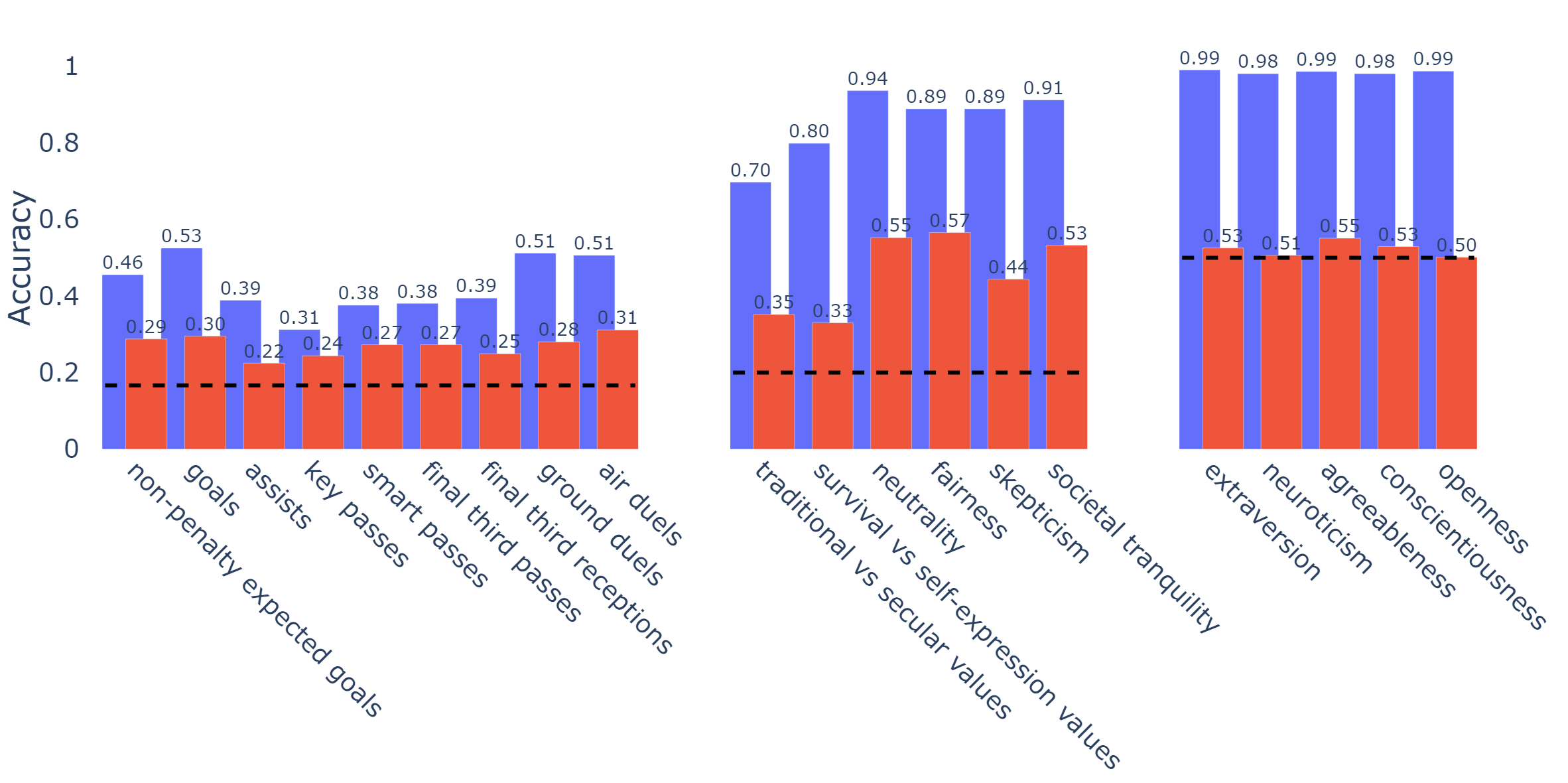}
   \caption{Comparison of the class labels generated by the normative model with classes reconstructed from the wordalisation. For each application, football scout (left), international survey (middle) and personality test (right), multiple wordalisations were generated for each data point, so that at least $10$ valid reconstructions per data point were found, and the mean accuracy is taken over all wordalistations. We compare the accuracy of the model for two different prompts, one in which data in the form of synthetic text, generate according to the normative model, was given (blue) and one in which the data was omitted (red). The dashed line indicates the expected accuracy if the class labels were randomly chosen according to a uniform probability distribution and lie at an accuracy of $\frac{1}{6}$, $\frac{1}{5}$ and $\frac{1}{2}$ for each application respectively.}
   \label{fig:performance}
\end{figure*}

 For the personality test and international survey applications, the high accuracy scores achieved when the normative model was used, shown in blue in Fig.~\ref{fig:performance}, indicate that the wordalisation were very faithful to the statistical descriptions provided by the normative model. For all applications, the test accuracies, where a description according to the normative model was provided, were higher across all factors compared to the control, where statistical data was omitted. This performance difference is further evidence for the influence of the statistical description generated by the normative model on the wordalisation. For comparison, we also indicate, using dashed lines in Fig.~\ref{fig:performance}, the accuracy expected in the case where each factor label was guessed, that is, selected at random with each label having equal probability of selection. For the football scout and international survey applications, the performance of the control was better than guessing across all factors, while the performance of the control for the personality test application was comparable to guessing. A likely explanation for this performance difference is the presence of relevant information in the LLM training data, which could plausibly be the case for football players and countries but not for anonymised individuals who took a personality test.


Fig.~\ref{fig:performance} also shows that the overall performance of the football scout application is lower than that of the other two applications. This aligns with the idea that more complex wordalisation statements like “W. Bony is a forward who has struggled to make a significant impact in the attacking phase of play” condense multiple observations from the synthetic text, making it harder for the LLM to accurately reconstruct the underlying synthetic text. In contrast, simpler statements in the personality test and international survey applications, such as “Andorra exhibits a notably secular stance, with citizens placing low importance on religious faith when considering the qualities they wish to instill in their children”, map more directly to the source data, facilitating more accurate LLM outputs. For this reason it was more difficult for the LLM to reconstruct the statistical descriptions of a football player from a wordalisation, compared to a person or a country.

\subsection{Human-as-a-judge}

For the human evaluation, we compare the {\it wordalisation} method against the \textit{control} and \textit{statistical} methods, following the same generation procedure as in the LLM-as-a-judge framework. In the {\it control} method the model is asked to describe an entity without access to numerical data. In the international survey and football scout example, the LLM used its training data to write descriptions about the country or football player, while for personality tests no fall back data was available. For the {\it statistical} method, the model is provided with z-scores and asked to generate a description based on these values. Using these three methods, we regenerate description for $30$ entities - $10$ from each example- resulting in  total of $90$ texts. Human raters were recruited through social media posts to go to an app (\url{https://wordalisations-evaluation.streamlit.app/}) where they were presented with a single entity at a time, along with its corresponding visualisation and a description from either the control, the statistical or the wordalisation treatment. To prevent direct comparisons and learning effects, each rater saw only one strategy per entity and never evaluates multiple descriptions of the same entity. Raters may evaluate as many descriptions as they wish and may stop at any point. For each description, we asked raters to assess the faithfulness of the description to the visualisation; its engagement; its usefulness for understanding or decision-making; and whether it contains hallucinations, defined as unsupported or fabricated claims. Responses are collected using ordinal Likert-style scales, with optional free-text comments. The evaluation interface presents one description per page, and a new entity–method pair is sampled immediately after submission. No personally identifiable information is collected; participants are informed of the study purpose and their right to withdraw at any time. We had 121 responses (from 26 unique individuals) with 42 texts from the control prompt, 40 texts from the statistical prompt and 40 texts form the wordalisation prompt. 

\begin{figure*}
      \centering
      \includegraphics[width=0.9\linewidth]{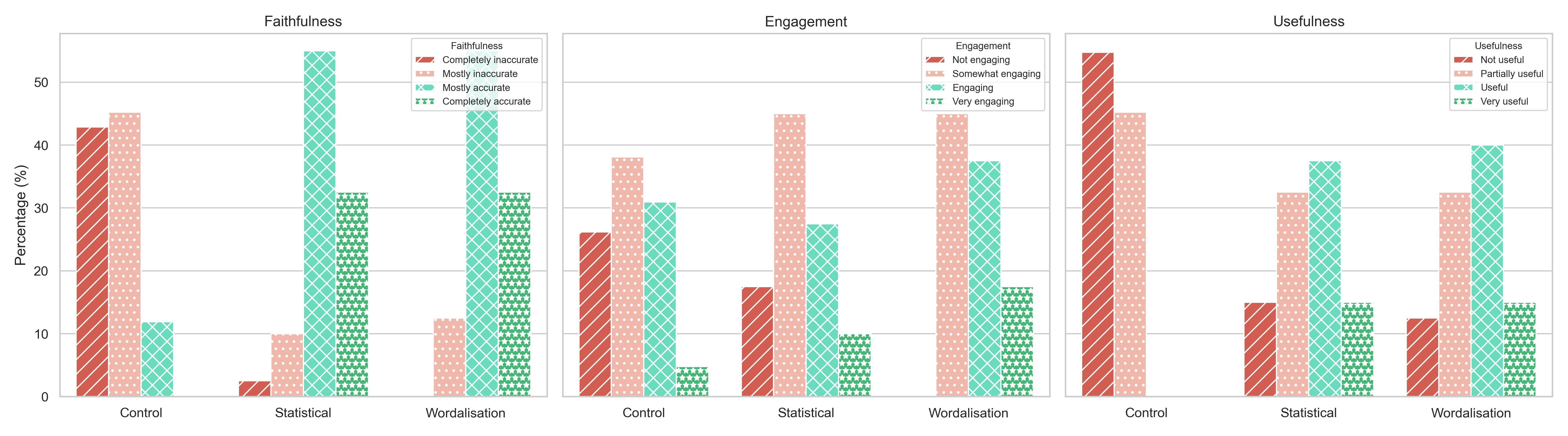}
   \caption{Human evaluation of three text variants (control, statistical, and wordalisation) across three metrics: faithfulness, engagement, and usefulness. Percentages of categorical ratings are shown for each text variant.}
   \label{fig:evaluation}
\end{figure*}

Fig.~\ref{fig:evaluation} shows how the wordalisation and statistical vastly texts outperformed the statistical text in terms of faithfulness and usefulness ($\chi^2=71.00$ for faithfulness and $\chi^2=40.50$ for engagement, both $p < 0.001$). Engagement was also significantly different when compared across all methods ($\chi^2=13.93$, $p = 0.03$). Importantly, when comparing only the statistical and wordalisation texts, we see that while they do not differ in terms of faithfulness ($\chi^2=1.11$, $p=0.77$) or usefulness ($\chi^2=0.12$, $p = 0.99$), there was a significant difference in engagement ($\chi^2= 8.43$, $p = 0.04$), with wordalisation preferred by the participants over statistical descriptions. 


\subsection{Documentation}

Following the model cards framework~\cite{mitchell2019model,crisan2022interactive,gebru2021datasheets,crawford2021atlas}, we emphasise the importance of clearly stating the model we are imposing on the data when creating the wordalisation: detailing how numerical values are translated into words, incorporating background information into prompts for the LLMs, and documenting the limitations of the wordalisation. This approach provides transparency on the risks of applying wordalisation to a new domain without performing extensive quantitative analysis.  Model cards offer a flexible framework for innovative thinking around best practices in modelling~\cite{tang2024ai}.
We documented each of our applications in the model card style adapted from~\cite{mitchell2019model} and can be found here\footnote{\url{https://github.com/amandinecaut/wordalisation/blob/prompt-card/Prompt\%20Cards/README.md}}.

All code developed for this work has been made publicly available in a public repository here\footnote{\url{https://github.com/amandinecaut/wordalisation}}. This includes the full implementation for the 3 applications and the evaluation pipeline used for assessment. By releasing the application and evaluation code, we aim to support transparency, reproducibility and broader adoption of the methodology.

\section{Discussion}

The key to the wordalisation approach is the combination of both statistical and normative models to represent data in words. It is this which leads us to our four steps of telling an LLM `who it is', `what it knows', `what data to use' and `how to answer' as a way to produce convincing texts. The `what data to use' step defines the statistical model of how we interpret the numbers in terms of z-scores and then a normative model provides our interpretation of these values. The other three steps give a text-based model of how we should interpret those scores, giving further background to our normative model. While z-scores are a relatively simple statistical model, we envisage (and are currently implementing) more advanced statistical models --- such as regression, principal component analysis and moving averages --- as the basis for wordalisation. 

Instead of fine-tuning a model, we force the pre-trained LLM to produce the texts which reflect our own model of that data. Previous work has shown that extensive fine-tuning of LLMs often does not significantly outperform (and in some cases underperforms) methods based on prefixed prompts on benchmark tasks~\cite{li2021prefix,jia2022visual}. Indeed, prompt-based or in-context learning~\cite{radford2019language,liu2023pre,brown2020language}, of which our wordalisations are examples, are a powerful way of utilising the power of LLMs in a way which respects their limitations. Our four-step automated method underlines just how well prompting can perform in application areas.

In human evaluations, we found that wordalisations outperformed a purely statistical description of the data in terms of engagement, while maintaining faithful and useful descriptions of the underlying data. This result contrasted with LLM evaluations, where the wordalisations for the football scout application, in particular, were unable to reconstruct the exact normative description. What human evaluation identified is that rather than containing inaccuracies, they simply  contained more varied language and were less formulaic. We thus see our human evaluations as highlighting the limitation of the LLM-as-a-judge approach, rather than as a benchmark test of performance. 

Ultimately, we do not envisage quantitative testing, by humans or machines, as the way to assess wordalisations. The wordalisation task involves generating text from structured data, making it a fundamentally different objective than captured by established benchmarks for Natural Language Processing (NLP) tasks. 
 For this reason, we have also adopted an approach that builds on model cards ~\cite{mitchell2019model}, emphasising the quality of documentation of the normative and statistical models we use, and the consequences of using these models, over output testing. As language models become integrated into an increasing number of tools, it is essential to provide transparent documentation of the thought process behind the prompt engineering~\cite{liu2023pre}. In future work, we will give a more complete outline of the model card approach to prompting.

Wordalisations, similar to visualisations, are powerful tools for data communication. They have the potential to make all types of data more accessible and comprehensible to a wider audience. However, evaluating them is inherently subjective, as absolute accuracy is often unattainable~\cite{tong2021inapplicability, ieee_chatgpt}. In our work, we have embraced this reality by prioritising transparency and honesty, akin to practices in visualisation. Rather than solely emphasising numerical evaluations, we advocate for providing thoughtful documentation through model cards and communicating model capabilities and limitations. Benchmarking and evaluation, while having some value, should support this process rather than be the sole target. We can compare this process with journalism: journalists rarely focus on benchmarking their work, but attempt to maintain high standards \cite{mintzer-sweeney_muth_2024,cairo2016truthful}. It is also consistent with the idea of context engineering \cite{context_engineering}: we are no longer at the stage of benchmark accuracy tests, but looking for ways to develop best practices for working with LLMs. Those exploring prompting should focus on documenting and standardising their methodologies to build a robust framework for future use. Looking ahead, prompting will play an increasingly significant role in software development and having a clear structure for documentation will be essential. 

\section*{Limitations}
A current limitation of the wordalisation method is its exclusive reliance on raw numerical data; it does not yet support the integration of other kind of data (such that categorical variables, time series). The current code does not yet include additional statistical models; however, extending the implementation to incorporate them is part of our future work.
The examples presented is intended as an illustration of the methods. The primary use case of this wordalisation is educational. It shows how to convert a dataframe of raw data into a text. A secondary use case might be for users to understand more about the Big Five personality test, World Value Survey, or Football player datasets. This version is not suitable for professional use, such as by a recruiter. However, the methods used here do have the potential to be adopted to a professional setting.


\bibliography{reference_nature}

\end{document}